\documentclass[10pt,aps,prb,twocolumn,amsmath,amssymb,superscriptaddress]{revtex4-1}
\usepackage{amsmath}
\usepackage{graphicx}
\usepackage{color}
\usepackage{multirow}
\usepackage[caption = false]{subfig}
\usepackage{array}

\def\N{\mathbf{n}}
\def\M{\mathbf{m}}
\def\K{\mathbf{k}}
\def\P{\mathbf{p}}
\def\G{\mathbf{g}}

\begin{document}
\renewcommand{\arraystretch}{1.2}

\title{Improved speed and scaling in orbital space variational monte carlo}
\author{Iliya Sabzevari}
\affiliation{Department of Chemistry and Biochemistry, University of Colorado Boulder, Boulder, CO 80302, USA}
\author{Sandeep Sharma}
\email{sanshar@gmail.com}
\affiliation{Department of Chemistry and Biochemistry, University of Colorado Boulder, Boulder, CO 80302, USA}
\begin{abstract}
In this work, we introduce three algorithmic improvements to reduce the cost and improve the scaling of orbital space variational Monte Carlo (VMC). First, we show that by appropriately screening the one- and two-electron integrals of the Hamiltonian one can improve the efficiency of the algorithm by several orders of magnitude. This improved efficiency comes with the added benefit that the resulting algorithm scales as the second power of the system size $O(N^2)$, down from the fourth power $O(N^4)$. Using numerical results, we demonstrate that the practical scaling obtained is in fact $O(N^{1.5})$ for a chain of Hydrogen atoms, and $O(N^{1.2})$ for the Hubbard model. Second, we introduce the use of the rejection-free continuous time Monte Carlo (CTMC) to sample the determinants. CTMC is usually prohibitively expensive because of the need to calculate a large number of intermediates. Here, we take advantage of the fact that these intermediates are already calculated during the evaluation of the local energy and consequently, just by storing them one can use the CTCM algorithm with virtually no overhead. Third, we show that by using the adaptive stochastic gradient descent algorithm called AMSGrad one can optimize the wavefunction energies robustly and efficiently. The combination of these three improvements allows us to calculate the ground state energy of a chain of 160 hydrogen atoms using a wavefunction containing $\sim 2\times 10^5$ variational parameters with an accuracy of 1 $mE_h$/particle at a cost of just 25 CPU hours, which when split over 2 nodes of 24 processors each amounts to only about half hour of wall time. This low cost coupled with embarrassing parallelizability of the VMC algorithm and great freedom in the forms of usable wavefunctions, represents a highly effective method for calculating the electronic structure of model and \emph{ab initio} systems.
\end{abstract}
\maketitle

Quantum Monte Carlo (QMC) is one of the most powerful and versatile tools for solving the electronic structure problem and has been successfully used in a wide range of problems\cite{RevModPhys.73.33,NigUmr-BOOK-99, TouAssUmr-AQC-15,KolMit-RPP-11,sorellabook}.
%Grossman95a,KentTowlerNeedsRajagopal00,FahyWangLouie88,Williamson98,BatHeyHenUbeMarSchUmrWil-PRB-06,NekoveeFoulkesNeeds01,FilUmr-JCP-96,UmrTouFilSorHen-PRL-07,Sorella07,Chang2008,White1989}. 
QMC can be broadly classified into two categories of algorithms, variational Monte Carlo (VMC)\cite{Mcmillan,CepCheKal-PRB-77} and projector Monte Carlo (PMC). In VMC one is interested in minimizing the energy of a suitably chosen wavefunction ansatz. The accuracy of VMC is limited by the functional form and the flexibility of the wavefunction employed. PMC on the other hand is potentially exact and several variants exist, such as diffusion Monte Carlo (DMC)\cite{GRIMM1971134,Anderson75,CeperleyAlder80}, Auxiliary field quantum Monte Carlo (AFQMC)\cite{ZhaKra-PRL-03,Mario-WIREs}
, Green's function Monte Carlo (GFMC)\cite{Runge1992,Trivedi1990,Sorella98} and full configuration interaction quantum Monte Carlo (FCIQMC)\cite{BooThoAla-JCP-09,CleBooAla-JCP-10,PetHolChaNigUmr-PRL-12,Holmes2016}. Although exact in principle, in practice all versions of PMC suffer from the fermionic sign problem when used for electronic problems, except in a few special cases. A commonly used technique for overcoming the fermionic sign problem is to employ the fixed-node or fixed-phase approximation that stabilizes the Monte Carlo simulation by eliminating the sign problem at a cost of introducing a bias in the calculated energies. The fixed-node or the fixed-phase is often obtained from a VMC calculation and thus the accuracy of the PMC calculation depends to a large extent on the quality of the VMC wavefunction. Thus VMC wavefunction plays a pivotal role and to a large extent determines the final accuracy obtained from a QMC calculation. (FCIQMC is unique in its ability to control the sign problem self-consistently by making use of the initiator approximation.) Traditionally the most commonly used variant of QMC has been the real space VMC followed by the fixed-node DMC calculation which is able to deliver results at the complete basis set limit. Although extremely powerful, one of the shortcomings of real space QMC methods is that there is less error cancellation than is typically observed in basis set quantum chemistry. This fact, coupled with the recent success of both AFQMC and FCIQMC (both of which work in a finite basis set), has led to renewed interest in development of new QMC algorithms that work in a finite basis set\cite{PhysRevB.96.085103,Tahara2008,Eric2013,PhysRevB.97.235103,PhysRevLett.118.176403,Sorella07,Neuscamman2016,PhysRevB.92.035122}. The present work is an attempt in this direction and presents algorithmic improvements for performing orbital space VMC calculations. 

To put the orbital space VMC in the broader context of wavefunction methods, it is useful to classify the various wavefunctions used in electronic structure theory into three classes. The wavefunctions in the first class allow one to calculate the expectation value of the energy, $\langle\psi|H|\psi\rangle$, deterministically with a polynomial cost in the number of parameters, examples of these include the configuration interaction wavefunction\cite{knowles1984new} and matrix product states\cite{White1992,White1993,Ostlund1995}. The second class of wavefunctions only allow one to calculate the overlap of the wavefunction with a determinant (or a real space configuration), $\langle \N|\psi\rangle$, deterministically with a polynomial cost in the number of parameters, examples include the Slater-Jastrow (SJ)\cite{PhysRevLett.10.159}, correlator product states\cite{Changlani2009}, Jastrow-antisymmetric geminal power (JAGP)\cite{Neuscamman2011}. Finally, the third class of wavefunction are the most general and do not allow polynomial cost evaluation of either the expectation value of energy or an overlap with a determinant, examples of these include the coupled cluster wavefunction\cite{RevModPhys.79.291} and the projected entangled pair states\cite{peps}. The first class of wavefunction are the most restrictive, but are the easiest to work with and often efficient deterministic algorithms are available to variationally minimize their energy. For the second class of wavefunctions one has to resort to VMC algorithms to both evaluate the energy and optimize the wavefunctions. Finally, the third type of wavefunctions are the most general and are often quite accurate but one needs to resort to approximations to evaluate their energies. In this work we will focus on the second class of wavefunction, in particular the wavefunction comprising of a product of the correlator product states and a Slater determinant (CPS-Slater). %We will show that the VMC procedure used to evaluate and optimize their energy can be made significantly faster while incurring only a small error in the calculated energy. 

The rest of the paper is organized as follows. In Section~\ref{sec:algo}, we will briefly outline the main steps of the VMC algorithm, followed by the new algorithmic innovation introduced to improve the efficiency and reduce the scaling of the algorithm. Next, in Section~\ref{sec:cps}, we will describe how the algorithm is used to calculate the energy and optimize the CPS-Slater wavefunction. We will also discuss in detail the cost and computational scaling of the algorithm%\footnote{During the preparation of this manuscript, we became aware of another article that describes a reduced scaling algorithm for orbital space VMC appeared on Arxiv (https://arxiv.org/pdf/1806.08778.pdf). The two algorithms are completely different and we expect the current algorithm to be significantly more efficient.}
. Finally in Section~\ref{sec:result} we will present the results obtained by this algorithm for model systems including the 1-D hydrogen chain and 2-D Hubbard model. 

\section{Algorithmic improvements}\label{sec:algo}
In VMC, the expectation value of the Hamiltonian for a wavefunction $\Psi(\P)$, where $\P$ are the set of wavefunction parameters, is calculated by performing a Monte Carlo integration.
\begin{align}
E =& \frac{\langle \Psi(\P) |H|\Psi(\P)\rangle}{\langle \Psi(\P) |\Psi(\P)\rangle} = \frac{\sum_{\N} |\langle \Psi(\P) |\N \rangle|^2 \frac{\langle \N|H|\Psi(\P)\rangle}{\langle \N|\Psi(\P)\rangle}}{\sum_{\N} |\langle \Psi(\P)|\N\rangle|^2} \nonumber\\
 =& \sum_{\N} \rho_{\N} E_L[\N] \nonumber\\
=& \langle E_L[\N]\rangle_{\rho_{\N}} 
\label{eq:vmc}
\end{align}
where, $E_L[\N] = \frac{\langle \N|H|\Psi(\P)\rangle}{\langle \N|\Psi(\P)\rangle}$ is the local energy of a determinant
$\N$, the last expression denotes that the energy is calculated by averaging the local energy over a set of determinants $\N$
that are sampled from the probability distribution $\rho_{\N} = \frac{ |\langle \Psi(\P) |\N\rangle|^2 }{\sum_{\K} |\langle \Psi(\P)|\K \rangle|^2}$. 
With Equation~\ref{eq:vmc} as the background, we can describe the VMC algorithm by splitting it into three main tasks
\begin{enumerate}
\item For a given determinant $|\N\rangle$ we have to calculate the local energy $E_L[\N]$.
\item We have to generate a set of determinants $|\N\rangle$ with a probability distribution $\rho_{\N}$ for a given wavefunction $\Psi$.
\item And finally, we need an optimization algorithm that can minimize the energy of the wavefunction by varying the parameters $\P$. 
\end{enumerate}
We introduce algorithmic improvements in each of these tasks which are described in detail in the next three sections.%after the define the MSJ parametrization.

% *** WRITE DOWN THE MSJ PARAMETRIZATION****
 
 \subsection{Reduced scaling evaluation of the local energy} The local energy $E_L[\N]$ of a determinant $|\N\rangle$ is calculated as follows
 \begin{align}
 E_L[\N] =& \frac{\langle \N|H|\Psi(\P)\rangle}{\langle \N|\Psi(\P)\rangle}\\
=&  \frac{\sum_{\M} \langle \N|H|\M\rangle \langle\M|\Psi(\P)\rangle}{\langle \N|\Psi(\P)\rangle}\\
=& \sum_{\M} H_{\N,\M} \frac{ \langle\M|\Psi(\P)\rangle}{\langle \N|\Psi(\P)\rangle},
 \end{align}
 where the sum is over all determinants $\M$ that have a nonzero transition matrix element ($H_{\N,\M} = \langle \N|H|\M\rangle$) with $\N$. The number of such non-zero matrix elements $H_{\N,\M}$ are on the order of $n_e^2 n_o^2$, where $n_e$ is the number of electrons and $n_o$ is the number of open orbitals. This number increases as a fourth power of the system size and a naive use of this formula results in an algorithm that scales poorly with the size of the problem. 
 
 To reduce the cost of calculating the local energy we truncate the summation over all $\M$ to just a summation over those $\M$, that have a Hamiltonian transition matrix element $H_{\N,\M} > \epsilon$, 
 \begin{align}
 E_L[\N, \epsilon] =&  \sum_{\M}^\epsilon H_{\N,\M} \frac{ \langle\M|\Psi(\P)\rangle}{\langle \N|\Psi(\P)\rangle},\label{eloc:hci}
 \end{align}
 where $\epsilon$ is a user defined parameter. Note that in the limit that $\epsilon \rightarrow 0$, we recover the exact local energy, $E_L[\N, \epsilon]  \rightarrow E_L[\N]$. It is useful to note that when a local basis set is used the number of elements $H_{\N,\M}$ that have a magnitude larger than a fixed non-zero $\epsilon$ scale quadratically with the size of the system. To see this, let's consider double excitations, the argument for the single excitations is similar. For a given determinant ($\N$), we can obtain another determinant by a double excitation of electrons from say orbitals $i$ and $j$ to orbitals $a$ and $b$ to give another determinant ($\M$) with a matrix element 
 $|H_{\N,\M}| = |\langle a b|i j\rangle - \langle a b | j i\rangle|$. Because of the locality of orbitals, all orbitals decay exponentially fast and thus the integral $\langle a b|i j\rangle$ is negligible unless $a$ is close to $i$ and $b$ is close to $j$. In other words this integral is non-zero for only O(1) instances of $a$ and $b$ and because there are an $O(N^2)$ number of possible $i$ and $j$, the total number of non-negligible integrals is also $O(N^2)$. Thus if we are able to efficiently screen the transition matrix elements for a given $\epsilon \neq 0$, no matter how small the $\epsilon$ is, we are guaranteed to obtain a quadratically scaling evaluation of the local energy. 
 
 To see how the parameter $\epsilon$ can be effectively used to dramatically reduces the cost of the local energy evaluation, let us examine the one electron and the two electron excitations separately. 
 
 First let's see how the more numerous two electron excitations can be screened. The value of the Hamiltonian transition matrix element between a determinant, $\M$, contributing to the summation, that is related to determinant, $\N$, by the double excitation of electrons from orbitals $i$ and $j$ to orbitals $a$ and $b$ is equal to 
 $|H_{\N,\M}| = |\langle a b|i j\rangle - \langle a b | j i\rangle|$. The magnitude of the Hamiltonian matrix element only depends on the four orbitals that change their occupation and does not depend on the rest of the orbitals of the determinants $\N$ or $\M$. To use this fact efficiently, we stores the integrals in the heat bath format whereby, for all pairs of orbitals $i$ and $j$ we store the tuples $\{a,b,\langle a b|i j\rangle - \langle a b | j i\rangle\}$ in descending order by the value of $|\langle a b|i j\rangle - \langle a b | j i\rangle|$. Now for generating all determinants $\M$ that are connected to $\N$ by double excitation that is larger than $\epsilon$, one simply loops over all pairs of occupied orbitals $i$ and $j$ in determinant $\N$. For each of these pairs of orbitals one loops over the tuple $\{a,b,\langle a b|i j\rangle - \langle a b | j i\rangle\}$, until the value of $|\langle a b|i j\rangle - \langle a b | j i\rangle| > \epsilon$ and one exits this inner loop as soon as this inequality is violated. The CPU cost of storing the integrals in the heat bath format is $O(N^4\ln(N^2))$, but this is essentially the same cost as reading the integrals $O(N^4)$ and only needs to be done once for the entire calculation. The algorithm described here is essentially exactly the same as the one used to perform efficient screening of two electron integrals in the Heat Bath Configuration interaction algorithm\cite{Holmes2016b}, which is a variant of the selected configuration interaction algorithm\cite{Huron1973} and is much more efficient than other variants.%If the orbitals $a$ and $b$ are unoccupied in determinant $\N$, then the determinant $\M$ is generated by exciting electrons from orbitals $i$ and $j$ to orbitals $a$ and $b$ and the contribution of the determinant $\M$ to the total local energy is evaluated as shown in Equation~\ref{eloc:hci}. 
 
It is important to recall that although it might seem that there are only $O(N^2)$ single excitations, and thus one does not need to screen them, this is in fact not true. The reason being the Hamiltonian transition matrix element between two determinants, $\N$ and $\M$, that are connected by the single excitation of an electron from orbital $i$ to orbital $a$ is given by $H_{\N,\M} = \langle i|a\rangle + \sum_{j\in occ} \langle ij|aj\rangle - \langle ij|ja\rangle$. There are $O(N^2)$ such connections and the cost of calculating each of these matrix elements scales as $O(N)$ thus making the entire cost $O(N^3)$. We have observed that without appropriate screening of these one electron integrals their cost starts to dominate the cost of calculating the local energy.  To screen the singly excited determinants we first calculate the quantity $S_{i,a} = |\langle i|a\rangle| + \sum_j |\langle ij|aj\rangle - \langle ij|ja\rangle|$, for all pairs of orbitals $i$ and $a$. Notice, that $S_{i,a} > |H_{\N,\M}|$. Thus, while generating singly excited determinants from a determinant $\N$, by exciting an electron from an occupied orbital $i$ to an empty orbital $a$, we discard all excitations where $S_{i,a} <\epsilon$.
 
 Thus for a given $\epsilon$, \emph{only determinants, $\M$, that make a non-zero contribution to the local energy are ever considered}. As we will demonstrate in Table~\ref{tab:eps}, this algorithm can be used to discard a very large fraction of the determinants from the summation of Equation~\ref{eloc:hci} providing orders of magnitude improvement in the efficiency of the algorithm while incurring only a small error in the overall energy. In Section~\ref{sec:eloc} we will show that the contribution to the local energy ($H_{\N,\M} \frac{ \langle\M|\Psi(\P)\rangle}{\langle \N|\Psi(\P)\rangle}$) for each of these $O(N^2)$ determinants $\M$ can be evaluated at a cost $O(1)$, allowing one to evaluate the local energy with a cost that scales quadratically with the size of the system. Recently, two other algorithms have appeared in literature that sample the matrix elements $H_{\N,\M}$ stochastically\cite{PhysRevLett.118.176403} and semistochastically\cite{ericScaling} instead of screening the summation deterministically as we have proposed here. Both these algorithms should have the same asymptotic scaling as the current algorithm, however, in practice the benefits of reduced scaling only manifest themselves for large system sizes.

\subsection{Continuous time Monte Carlo for sampling determinants}
The usual procedure for generating determinants $\N$ according to a probability distribution $\rho_{\N}$ uses the Metropolis-Hastings algorithm in which a Markov chain is realized by performing a random walk according to the following steps 
\begin{enumerate}
\item Starting from a determinant $\N$, a new candidate determinant $|\M\rangle$ is generate with a proposal probability $P(\M \leftarrow \N)$.
\item The candidate determinant is then accepted if a randomly generated number between 0 and 1 is less than $\min\left(1, \frac{\rho(\M) P(\N \leftarrow \M)}{\rho(\N)P(\M \leftarrow \N)}\right)$, otherwise we stay at determinant $\N$ for another iteration.		
\end{enumerate}
The algorithm guarantees that with sufficiently large number of steps one generates determinants according to the probability distribution $\rho$ as long as the principle of ergodicity is satisfied. This states that it should be possible to reach any determinant $\K$ from a determinant $\N$ in a finite number of steps. The draw back of the algorithm is that if the one chooses the proposal probability distribution poorly, such that determinants $\M$ that have a small $\rho(\M)$ are proposed with a high probability, then several moves will be rejected leading to long autocorrelation times. Although the algorithm places very few restriction on the proposal probability distribution, and one can in principle devise a probability distribution that leads to very few rejections, in practice it is far from easy to do this. %To the best of authors knowledge a systematic study of proposal probability distribution that works well for both the strongly and weakly correlated systems is not available.

In this work we bypass the need for devising complicated proposal probability distributions, by using the continuous time Monte Carlo (CTMC), also known as the kinetic Monte Carlo (KMC), Bortz-Kalos-Lebowitz (BKL) algorithm\cite{BORTZ197510} and Gillespie's algorithm\cite{GILLESPIE1976403}. This is an alternative to the Metropolis-Hastings algorithm and has the advantage of being a rejection free algorithm and every proposed move is accepted. Just as in the case of the Metropolis-Hastings algorithm, there is considerble flexibility in how the algorithm is implemented but in this work we use the following steps:
\begin{enumerate}
\item Starting from a determinant $\N$ calculate the quantity 
\begin{align}
r(\M\leftarrow \N) = \left(\frac{\rho(\M)}{\rho(\N)}\right)^{1/2} = \left|\frac{ \langle\M|\Psi(\P)\rangle}{\langle \N|\Psi(\P)\rangle}\right| \label{eq:rate}
\end{align}
 for all determinants $\M$ that are connected to $\N$ by a single excitation or a double excitation. %(with a Hamiltonian transition matrix element with an absolute value greater than $\epsilon$).
\item Calculate the residence time $t_\N = \frac{1}{\sum_{\M} r(\M\leftarrow \N) }$ and stay on the walker $\N$ for the time $t_\N$ (the residence time can also be viewed as a weight).
\item Next, a new determinant is selected, without rejection, out of all the determinant $\M$ with a probability proportional to $r(\M\leftarrow \N)$.		
\end{enumerate}

To get a better understanding of the algorithm, it is useful to think of each determinant ($\N$) as a reactive species in a unimolecular reaction network that can be transformed into another determinant ($\M$) through a unimolecular reaction with the ratio of the forward and reverse rates determined by the equilibrium constant (in our case $\frac{k(\M\leftarrow \N)}{k(\N\leftarrow \M)} = K_{eq} = \left | \frac{\langle \M|\Psi(\P)\rangle}{\langle \N|\Psi(\P)\rangle} \right |^2$). As long as all determinants are participating in the reaction network, it will reach an equilibrium state in which a determinant ($\N$) will have a concentration   
\begin{align}
c(\N) \propto \left|\langle \N|\Psi(\P)\rangle\right | ^2.
\end{align}
The CTMC algorithm can be viewed as a way of driving this unimolecular reaction network stochastically to equilibrium. 

Notice, that the same equilibrium state will be reached irrespective of the relative magnitude of the rate constants $k(\M\leftarrow \N)$ versus $k(\K\leftarrow \N)$.
%{\color{red} I think you mean irrespective of the magnitude of the the rate constants provided that the ratio is unchanged.  Also, K should be M.}
Within the reaction network picture, this is akin to catalyzing a reaction which can lead to faster equilibration but does not change the equilibrium state itself. This allows considerable freedom in how the algorithm is implemented. The rates chosen by us in Equation~\ref{eq:rate} are only one example of a choice that one can make. In fact, as long as the algorithm satisfies the condition of detailed balance $\frac{r(\M\leftarrow \N)}{r(\N\leftarrow \M)} = \frac{\rho(\M)}{\rho(\N)}$, and the principle of ergodicity, it is guaranteed to sample the determinants with the correct probability.
%{\color{red} I would be inclined to say "a condition akin to detailed balance", but either is OK.  I am not aware of anyone who has used this for VMC.  You say it is "rarely used" but do you know of any example in VMC as opposed to PMC?  If you don't know any, then may be you should say that you are extending the CTCM algorithm to VMC.}

Although the CTMC can be used in virtually all Monte Carlo simulations, to the best of our knowledge it has never been used in VMC before. This is partly because one has to calculate and store the quantities $r(\M\leftarrow \N)$ for a potentially large set of connected configurations and this is usually quite expensive. However, it is interesting to note that in the VMC algorithm, the quantities $\left|\frac{ \langle\M|\Psi(\P)\rangle}{\langle \N|\Psi(\P)\rangle}\right|$ are already used in the calculation of the local energy (see Equation~\ref{eloc:hci}) and just by storing those quantities the CTMC algorithm can be used with almost no overhead. We notice that using the CTMC algorithm leads to a shorter autocorrelation time and results in a more efficient VMC calculation.

A potential difficulty that might arise due the use of CTMC is that one could encounter a situation in which although a set of determinants have a large overlap with the current wavefunction, they are never reached because we start our Monte Carlo simulation from a determinant that is not connected to these determinants through a set of significant Hamiltonian transition matrix elements (ergodicity of the simulation is broken). For example let us imagine a situation in which two states $|\Psi_1\rangle$ and $|\Psi_2\rangle$ are nearly degenerate but have different irreducible representations. If we introduce a small perturbation in the Hamiltonian that breaks the symmetry, then the resulting ground state will be a linear combination of the two states $|\Psi_1\rangle$ and $|\Psi_2\rangle$. In such a situation, the CTCM algorithm when performed with a screening parameter $\epsilon$, that is larger than the magnitude of the perturbation, will not be able to sample both states and will most likely be stuck in one or the other state depending on the initial determinant chosen to start the Monte Carlo calculation. This situation is difficult to diagnose because it is likely that the energy obtained will be reasonably accurate, but the properties such as correlation function, etc. will be quite inaccurate. Although it is difficult to be certain that such an eventuality is eliminated, it can be avoided to a large extent by fully localizing the orbital basis, and trying to break all possible spatial symmetries. This will ensure that the approximate symmetries of the problem will not influence the Monte Carlo walk.

\subsection{AMSGrad algorithm for optimizing the energy}
The optimized wavefunction ($\Psi(\P)$) is obtained by minimizing its energy with respect to its parameters $\P$. The gradient of the energy of the wavefunction with respect to  $\P$ can be evaluated as follows
\begin{align}
\G_i =& \frac{\partial E}{\partial \P_i}\\
=& \frac{\partial\langle \Psi(\P) |H|\Psi(\P)\rangle/\langle \Psi(\P) |\Psi(\P)\rangle}{\partial \P_i}\nonumber\\
=& 2\frac{\langle \Psi_i(\P) |H - E|\Psi\rangle}{\langle \Psi(\P) |\Psi(\P)\rangle}\nonumber\\
=& 2 \frac{\sum_{\N} |\langle \Psi(\P) |\N \rangle|^2  \frac{\langle \Psi_i(\P)|\N\rangle}{\langle \Psi(\P)|\N\rangle} \frac{\langle \N|H - E|\Psi(\P)\rangle}{\langle \N|\Psi(\P)\rangle}}{\sum_i |\langle \Psi(\P)|\N\rangle|^2}\nonumber\\
=&  2 \left\langle  \frac{\langle \Psi_i(\P)|\N\rangle}{\langle \Psi(\P)|\N\rangle} (E_L[\N]-E)\right\rangle_{\rho_{\N}} 
\label{eq:grad}
\end{align}
where, $|\Psi_i(\P)\rangle = \left|\frac{\partial\Psi(\P)}{\partial \P_i}\right\rangle$ and in going from the first line to the second we have assumed that all parameters are real. Note that during the calculation of the energy the determinants $\N$ are generated according to the probability distribution $\rho_{\N}$ and for each of these determinants a local energy $E_L[\N]$ is evaluated. Thus to obtain the gradient of the energy, the only additional quantity needed is a vector of gradient ratios $\frac{\langle \Psi_i(\P)|\N\rangle}{\langle \Psi(\P)|\N\rangle}$. We will show that these can be calculated at a cost that scales linearly with the number of parameters in the wavefunction. In the wavefunctions that we have used in this work, the number of parameters themselves scale quadratically with the size of the system. Thus the gradient can be calculated at a cost that scales no worse than the local energy.

The two most common algorithms for minimizing the energy of the wavefunctions in VMC are the linear method (LM)\cite{UmrTouFilSorHen-PRL-07,Toulouse2007,TouUmr-JCP-08} and the stochastic reconfiguration (SR)\cite{PhysRevB.64.024512}. The former is a second order method and can be viewed as an instance of the augmented Hessian method in which one repeatedly solves the generalized eigenvalue equation
\begin{align}
\left(\begin{array}{cc}
E_0 & \G ^T\\
\G & \overline{H}\end{array}\right) \left(\begin{array}{c}
1\\
\Delta \P\end{array}\right) =& \left(\begin{array}{cc}
1 & 0\\
0 & \overline{S}\end{array}\right) \left(\begin{array}{c}
1\\
\Delta \P\end{array}\right)
\end{align}
where, $\overline{S}_{ij} = \langle \overline{\Psi}_i|\overline{\Psi}_j\rangle$,  $\overline{H}_{ij} = \langle \overline{\Psi}_i|H|\overline{\Psi}_j\rangle$, $\Delta\P$ is the update to the wavefunction parameters and $|\overline{\Psi}_i\rangle = \frac{1}{\sqrt{\langle\Psi|\Psi\rangle}}\left(|\Psi_i\rangle - \frac{\langle\Psi|\Psi_i\rangle}{\langle\Psi|\Psi\rangle}|\Psi\rangle\right)$. (Here we won't go into the details of how the matrices $\overline{S}$ and $\overline{H}$ are calculated.)

The SR algorithm can be thought to be performing projector Monte Carlo by repeated application of the propagator $(I - \Delta t H)$ to the wavefunction $|\Psi(t)\rangle$ and at each step projecting the resulting wavefunction on to the tangent space constructed from the wavefunction gradients $|\Psi_i\rangle$. It can be shown that this results in a linear equation

\begin{align}
\overline{S}_{ij} \Delta\P = -(\Delta t)\G 
\end{align}
which has to be solved at each step to obtain $\Delta\P$, the update to the wavefunction parameters.
  Note that in both these methods, one has to solve an equation at each iteration which has a CPU cost of $O(N_p^3)$ and a memory cost of $O(N_p^2)$, where $N_p$ is the number of wavefunction parameters, which we will show scales quadratically with the size of the system. The CPU cost can be reduced to $O(n_{s} N_p$) and $O(n_{s} N^4$) respectively for SR and LM method by using a direct method which avoid building the matrices, where $n_{s}$ is the number of Monte Carlo samples used in each optimization iteration\cite{Neuscamman2012}.
%{\color{red} It seems to me the CPU cost may actually go up when using a direct method.}
  
  Both these methods are quite effective at optimizing the wavefunction, in particular, the LM requires fewer iterations and is often able to find the optimized wavefunction containing about 1000 parameters in less than 10 iterations. The effectiveness of the method is somewhat adversely affected while using direct method because one often needs to include level shifts to remove the singularity in the overlap matrix $\overline{S}$, which can result in slower convergence.% A possible drawback is that in both these methods the parameter update $\Delta \P$ depends non-linearly on the elements of matrices $\overline{S}$ and $\overline{H}$, which in turn contain stochastic errors. This implies that the calculated parameters will be systematically biased away from their optimal values and the bias will only go to zero in the limit of infinite number of Monte Carlo samples. %{\color{red} True, but is this ever important in practice?}
  
  In this work we instead use a flavor of the adaptive stochastic gradient descent (SGD) method called AMSGrad\cite{Reddi2017}. The use of SGD methods have become popular in machine learning, where one is often interested in optimizing a cost function that depends non-linearly on a set of parameters and which can only be evaluated with a stochastic error by using batches of finite samples. This problem is of course very similar to the one we are interested in solving in VMC. The use of these adaptive stochastic gradient descent in the context of VMC was first proposed by Booth and coworkers\cite{PhysRevLett.118.176403}.
  
  The stochastic gradient descent methods have the advantage that, (1) their CPU and memory cost scales linearly with the number of wavefunction parameters and (2) the update in parameters depends strictly linearly on the gradient and thus does not introduce a systematic bias. The disadvantage of these methods is that traditionally they have been thought to be too slow and needing several thousand iterations to reach convergence, thus rendering them relatively ineffective. We show that by choosing appropriate parameters in the AMSGrad algorithm, we obtain convergence in a few tens to hundreds of iterations, which coupled with the fact that each iteration is much cheaper than LM and SR algorithm, makes it a powerful algorithm to solve challenging non-linear optimization problems in VMC.

A general algorithm for adaptive methods is as follows
\begin{align}
\mathbf{m}^{(i)} =& f(\G^{(0)},\cdots, \G^{(i)})\\
\mathbf{n}^{(i)} =& g(\G^{(0)},\cdots, \G^{(i)})\\
\Delta\P_j =& -\alpha^{(i)} \mathbf{m}_j^{(i)}/\sqrt{\mathbf{n}^{(i)}_j}
\end{align}
where $f$ and $g$ are some functions that take in all the past gradients and generate vectors $\mathbf{m}^{(i)}$ and $\mathbf{n}^{(i)}$, that are then used to calculate the updates $\Delta\P$ to the parameters. The various flavors on adaptive methods, ADAGrad\cite{adagrad}, RMSProp\footnote{T. Tieleman and G. Hinton. RmsProp: Divide the gradient by a running average of its recent magnitude. COURSERA: Neural Networks for Machine Learning, 2012.}, ADAM\cite{adam} and AMSGrad\cite{Reddi2017} differ in the function $f$ and $g$. In AMSGrad $f$ and $g$ respectively calculate the exponentially decaying moving average of the first and second moment 
\begin{align}
\mathbf{m}^{(i)} =& (1-\beta_1)  \mathbf{m}^{(i-1)} + \beta_1\G^{(i)}\label{eq:m1}\\
\mathbf{n}^{(i)} =& \max(\mathbf{n}^{(i-1)}, (1-\beta_2) \mathbf{n}^{(i-1)} +  \beta_2(\G^{(i)}\cdot\G^{(i)})\label{eq:m2}
\end{align}
of the gradients ($\G$) respectively, with the caveat that the second moment at iteration $(i)$ is always greater than at iteration $(i-1)$, i.e. $\mathbf{n}^{(i)} \geq \mathbf{n}^{(i-1)}$. This ensures that the learning rate, $\alpha^{(i)}/\sqrt{\mathbf{n}^{(i)}}$, is a monotonically decreasing function of the number of iterations. This was shown to improve the convergence of AMSGrad for a synthetic problem over ADAM, which is essentially identical to AMSGrad with the only difference being that $g$ calculates the exponentially decaying moving average of the second moment of the gradient. In our experiments with various VMC calculations we have found AMSGrad to always be superior to ADAM and RMSProp. For most of the systems studied in this article, unless otherwise specified, we have used the parameters $\alpha^{(i)} = 0.01, \beta_1 = 0.1, \beta_2=0.01$, which are significantly more aggressive than the recommended values\cite{Reddi2017} of $\alpha^{(i)} = 0.001, \beta_1 = 0.1, \beta_2=0.001$.

\section{Computational considerations}\label{sec:cps}
In this section we will first briefly describe the wavefunction consisting of the product of a correlator product state\cite{Neuscamman2011,Neuscamman2012} and a Slater determinant and then show that the cost of each stochastic sample is dominated by operations that cost $O(N^2)$ for ab-initio Hamiltonians and $O(N)$ for the Hubbard model.

\subsection{Correlator product states and Slater determinant} 
The wavefunction ($|\Psi\rangle$) used in this work consists of a product of the correlator product states (CPS) ($\hat{C}$) and a Slater determinant ($|\Phi\rangle$)
\begin{align}
|\Psi\rangle =& \hat{C}|\Phi\rangle\\
|\Phi\rangle =& \prod_{x<N} \left(\sum_{i} \theta_{ix} a_i^\dag\right) |0\rangle\\
\hat{C} =& \prod_{\mathbf{\lambda}} \left( \sum_{\mathbf{n_\lambda}} c_{\mathbf{n_\lambda}} \hat{P}_{\mathbf{n_\lambda}}\right)
\end{align}
where, the subscripts ($i, j, \cdots$) represent the local orbital basis and subscripts ($x, y, \cdots$) represent the delocalized molecular orbitals, $N$ is the number of electrons, $\theta$ is the matrix of molecular coefficients, $\lambda$ represents the correlators (see Figure~\ref{fig:lattice}), $\mathbf{n_\lambda}$ represents all the Fock states in the correlator $\lambda$ and finally the $\hat{P}_{\mathbf{n_\lambda}} = |\mathbf{n_\lambda}\rangle \langle \mathbf{n_\lambda}|$ is the projector onto the Fock states of the correlator. CPS is an instance of the tensor network state, and is closely related to several other wavefunctions such as Huse-Elser wavefunction, the Jastrow factors used in the resonating valence bond states and Laughlin wavefunction. In all our calculations we use local correlators containing at most 5 sites. Although the number of parameters increase exponentially with the size of the correlator, the number of local correlators themselves only increase linearly with the size of the system. In this work we always use local correlators and thus the number of parameters in CPS $c_{\mathbf{n\lambda}}$ only increases linearly with the size of the system, however the number of parameters in the matrix $\theta$ increases quadratically. %{\color{red} I would change $\mathbf{n\lambda}$ to $\mathbf{n_\lambda}$ everywhere.}

\subsection{Local energy and Gradient ratios\label{sec:eloc}} At each optimization step in VMC, one needs to evaluate the local energy and the gradient of the energy with respect to the wavefunction parameters. To evaluate these, one needs to be able to calculate the overlap ratio $\frac{\langle \M|\Psi(\P)\rangle}{\langle \N|\Psi(\P)\rangle}$ for all the $O(N^2)$ determinants, $\M$, that are connected to the current determinant, $\N$, with a non-negligible Hamiltonian transition matrix element, as well as the gradient ratio $\frac{\langle \Psi_i(\P)|\N\rangle}{\langle \Psi(\P)|\N\rangle}$ for all the $O(N_p) = O(N^2)$ parameters in the wavefunction.

The ratio of the overlap is equal to the product of the ratios of the overlaps with the CPS and the Slater determinant
\begin{align}
\frac{\langle \M|\Psi(\P)\rangle}{\langle \N|\Psi(\P)\rangle} = \frac{\hat{C}[\M]}{\hat{C}[\N]}\frac{\det(\theta[\M])}{\det(\theta[\N])}  
\end{align}
where, $\theta[\M]$ is the square matrix constructed by only retaining those rows and columns from $\theta$ that are occupied in $\M$ and $\Phi$.

The CPS overlap with a determinant, $\hat{C}[\M]$, is just the product of the coefficients ($c_{\mathbf{n_\lambda}}$) of all correlators ($\lambda$) present in $\M$. Therefore, the ratio of the CPS overlap ($\hat{C}[\M]/\hat{C}[\N]$) is equal to the ratio of the coefficients for only those correlators that contain at least one of the four orbitals that differ between $\N$ and $\M$, as all correlators in common cancel. This ratio can be evaluated with an O(1) cost as follows. For each orbital in the system we maintain a list of all the correlators that contain that orbital. Because of the locality of the correlators, there are only a maximum of O(1) such local correlators per orbital and thus the ratio of the correlators can be calculated in a time that scales as O(1).

The cost of calculating the ratio of the Slater determinants ($\det(\theta[\M])/\det(\theta[\N])$) can be reduced from $O(N^3)$ to $O(N)$ by using the Woodbury lemma\footnote{M. Brookes, The Matrix Reference Manual, (2011); see online at http://www.ee.imperial.ac.uk/hp/staff/dmb/matrix/intro.html.} 
\begin{align}
\frac{\det(\theta[\M])}{\det(\theta[\N])} = R[\N]_{ai} = \sum_x \theta_{ax}(\theta[\N]^{-1})_{xi}  \label{eq:R}
\end{align}
which holds when $\M$ is obtained from $\N$ by exciting an electron from orbital $i$ to orbital $a$. Similar expressions exist for the ratio of the determinants that are related by a double excitation, for example, when two $\alpha$ electrons are excited from orbitals $i$ and $j$ to orbitals $a$ and $b$, then the ratio is given by $R[\N]_{a,i}R[\N]_{b,j}- R[\N]_{b,i}R[\N]_{a,j}$. Thus, by precalculating and storing the matrix $R[\N]_{a,i}$ the ratios of determinants that are related by single and double excitations can also be calculated with an $O(1)$ cost. To calculate $R$ one needs to have access to the inverse of the matrix $\theta[\N]$. Performing direct inversion is expensive and has an $O(N^3)$ cost. We avoid the expensive direct inversion by simply updating the inverse of the matrix using the Sherman-Morrison formula 
\begin{align}
\theta[\M]^{-1} = \theta[\N]^{-1} - \frac{ \left(\theta[\N]^{-1} u\right) \left(v^T\theta[\N]^{-1}\right)}{I + v^T\theta[\N]^{-1}u}
\end{align}
that has a cost $O(N^2)$, where $\M$ and $\N$ are related by a single excitation from orbital $i$ to orbital $a$, $u_{x1} =\delta_{xi}$ and $v_{x1} = \theta_{ax} - \theta_{ix}$. Thus one only needs to calculate the inverse once at the beginning of the calculation with an $O(N^3)$ cost and all subsequent updates can be calculated at an $O(N^2)$ cost. It is also worth noting that calculating the matrix $R[\M]$ itself has an $O(N^3)$ cost, however, this can be reduced to $O(N^2)$ by updating the matrix $R[\N]$ at each iteration
\begin{align}
R[\M] = R[\N] - \theta \frac{ \left(\theta[\N]^{-1} u\right) \left(v^T\theta[\N]^{-1}\right)}{I + v^T\theta[\N]^{-1}u}
\end{align}
which has a cost $O(N^2)$.

The elements of the vector of gradient ratios $\frac{\langle \Psi_i(\P)|\N\rangle}{\langle \Psi(\P)|\N\rangle}$ for CPS and Slater determinant parameters can be calculated as 
\begin{align}
\frac{\langle \Psi_{c_{\mathbf{n}\lambda}}(\P)|\N\rangle}{\langle \Psi(\P)|\N\rangle} =& \frac{\langle \Psi(\P)|\N\rangle}{c_{\mathbf{n}\lambda}}\\
\frac{\langle \Psi_{\theta_{xi}}(\P)|\N\rangle}{\langle \Psi(\P)|\N\rangle} =& \langle \Psi(\P)|\N\rangle\left(\theta[\N]^{-1}\right)_{ix},
\end{align} 
where each equation can be implemented in O(1) time as long as matrix $\theta[\N]^{-1}$ is available.

%The above arguments hold without any changes if our slater determinant consists of a product of slater determinants containing an $\alpha$-electron determinant and a $\beta$-electron determinant.

\subsection{Computational scaling of the algorithm} 
The Table~\ref{tab:cpu} contains the formal computational scaling of the various steps of the algorithm. The cost of the algorithm is $O(n_sN^2)$ down from the usual algorithm that has a cost of $O(n_sN^4)$. The essential reason for the reduction in the cost is the introduction of the screening parameter $\epsilon$, and the use of a first order gradient descent method. 

\begin{table}[htp]
\caption{Scaling of the cost of various steps per optimization step in the current and older VMC algorithms when a CPS-Slater wavefunction is used. In the table we have assumed that the number of parameters scales as the square of the size of the system ($N$) and $n_s$ is the number of stochastic samples.}\label{tab:cpu}
\begin{tabular}{llcc}
\hline
\hline
Step & ~~~~~&\multicolumn{2}{c}{Cost}\\
&&~Current~&~~Other~~\\
\hline
\multicolumn{4}{l}{\emph{Local Energy Calculation}} \\
Singles&~~&$O(n_sN^2)$&$O(n_sN^3)$\\
Doubles&~~&$O(n_sN^2)$&$O(n_sN^4)$\\
\\
\multicolumn{4}{l}{\emph{Update Determinant $\N \rightarrow \M$}} \\
Updating determinant inverse($\theta[\N]^{-1}$)&&$O(n_sN^2)$&$O(n_sN^2)$\\
Precalculate matrix $R$ (Eq.\ref{eq:R})&&$O(n_sN^2)$&$O(n_sN^3)$\\
\\
\multicolumn{3}{l}{\emph{Optimizing the wavefunction}} \\
AMSGrad& &$O(N^2)$& - \\
LM/SR& & - &$O(N^6)$\\
SR (direct)& & - &$O(n_sN^2)$\\
LM (direct)& & - &$O(n_sN^4)$\\
\hline
\end{tabular}
\end{table}

To reason about the scaling of $n_s$ with the system size, let us define the scaling of a method as the cost of performing a calculation to obtain a constant error per electron as the size of the system increases. This is in line with the usual definition used to define a linear scaling method in electronic structure theory. With this definition the scaling of the method is closely linked to the size-extensivity of a method, for example, with this definition the scaling of a method that is not size-extensive  is not a very useful concept because the error per particle keeps on increasing and in the limit of a large enough system size the correlation energy obtained goes to zero. Let us consider the wavefunction of a large system composed of independent, unentangled, and identical subunits. The CPS-Slater wavefunction where both the correlators and Slater determinant are optimized has enough flexibility to describe the wavefunction of such a system by factorizing the overall wavefunction into a product of wavefunctions of individual subunits. For such a system the variance ($\sigma^2$) of the total wavefunction scales linearly with the number of such subunits or the size of the systems i.e. $\sigma^2 \propto N$. The error estimate ($e$) of a Monte Carlo calculation that uses $n_s^i$ uncorrelated samples is thus $e \propto \left(N/n_s^i\right)^{1/2}$. Note that $n_s^i$ is not equal to the number of stochastic samples $n_s$, because in our calculations each stochastic update usually moves a single or at most two electrons. This indicates that there is likely a serial correlation of length $N$ in the Monte Carlo samples and thus $n_s \approx n_s^i N$. Thus $e \propto \left(N^2/n_s \right)^{1/2}$ and if we want a constant error per particle ($e/N$), then we need to perform a constant number of Monte Carlo iterations. In other words, $n_s$ needed is independent of the size of the system if a constant error per particle is needed.

The analysis of scaling becomes somewhat more complicated because often the optimization algorithms are quite sensitive to the noise in the calculated energy and gradient. Thus, one might envision a scenario in which given a wavefunction one can calculate an accurate relative energy with an $n_s$ that is independent of the system size, however, to optimize the wavefunction to minimize the energy a system size independent $n_s$ is not sufficient. In Section~\ref{sec:optimizer} we give numerical evidence using the hydrogen chain of increasing size to show that the adaptive SGD is stable with respect to the stochastic noise and displays only a weak system size dependent convergence rate with a constant $n_s$.
%{\color{red} I think it is too strong to say it is "system independent" when there clearly is some dependence on system size.}

\section{Results}\label{sec:result}
In this section we give more details about the saving in the CPU cost and error incurred due to the use of the screening parameter $\epsilon$; we will also discuss the effectiveness of the AMSGrad in optimizing the wavefunction and will end with results on benchmark systems.

\subsection{Accuracy and scaling}
Table~\ref{tab:eps} shows the effect on the accuracy and CPU cost as the parameters $\epsilon$ is varied. Notice that there is almost a factor of 300 reduction in the CPU cost accompanied by an error of 41 milliHartree as we go from $\epsilon=0$ to $\epsilon = 10^{-3}$. However, interestingly, the optimized wavefunction itself is quite accurate as is indicated by the energies in the final column entitled ``Final Energy". This energy is calculated by evaluating the energy using a tighter $\epsilon = 10^{-8}$ with a wavefunction that was obtained by optimizing using a looser $\epsilon$. Thus a useful strategy for calculating the energies is to perform the optimization with a loose $\epsilon$ and then calculate the final energy using a single shot calculation with a tighter $\epsilon$. It should be noted that the calculations in Table~\ref{tab:eps} were performed on a $H_{50}$ at a bond length of 2.2 $a_0$, and the improvements in CPU cost will be larger when one goes to larger systems or longer bond lengths. 

\begin{figure}
\begin{center}
\includegraphics[width=0.48\textwidth]{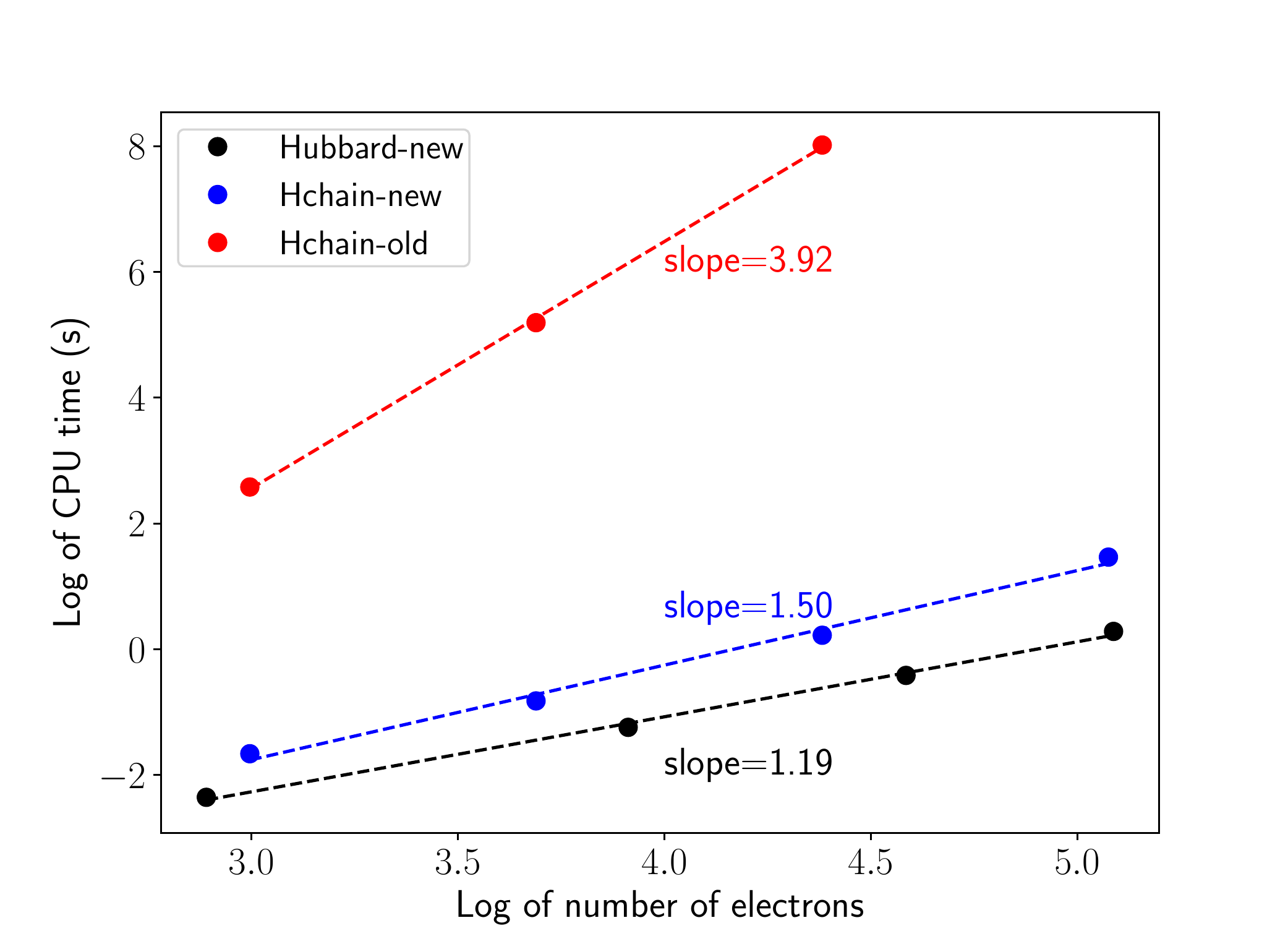}
\end{center}
\caption{The figure shows the scaling of the original and the new algorithm (scaling is equal to the slope) with the number of electrons for a 2-D Hubbard model and a chain of hydrogen atoms, when an $\epsilon=10^{-4}$ was used in the new algorithm. Hchain-new and Hchain-old respectively show the cost of the calculation with the current and original algorithm respectively on a hydrogen chain of varying lengths with a wavefunction consisting of 5-site local correlators and a slater determinant. The new algorithm is also used on the Hubbard model with the same wavefunction and shows that the scaling is closer to linear with the size of the system (see text for more discussion).}
\label{fig:scaling}
\end{figure}

\begin{table}
\caption{These calculations were performed on an open chain of 50 hydrogen atoms with a bond length of 2.2 $a_0$ with a minimal sto-6g bais set. The wavefunction consists of a product 5-site CPS with a Determinant obtained from a restricted Hartree Fock calculation. The CPU cost in the table is the time in seconds needed to perform 1000 stochastic samples. The optimized energy is obtained by minimizing the energy with respect to the CPS parameters, keeping the determinant fixed.  The final energy is obtained by using the optimized wavefunction for the given $\epsilon$ and then running a single point calculation with an $\epsilon=10^{-8}$.}\label{tab:eps}
\begin{tabular}{lccc}
\hline
\hline
$\epsilon$ & CPU & ~~Optimized~~  & ~~Final~~ \\
& cost (s) & Energy ($E_h$) & Energy ($E_h$)\\
\hline
0 &440.0 & - & -\\
$10^{-06}$&	4.9&-26.540 & -26.540\\
$10^{-05}$&	2.8 &-26.541 & -26.540\\
$10^{-04}$&	1.9 &-26.541 & -26.540\\
$10^{-03}$& 1.5	&-26.582 & -26.539\\
$10^{-02}$& 1.4	&-26.739 &-26.537\\
\hline
\end{tabular}
\end{table}
Next let us examine the scaling of the algorithm with the size of the system for a 1-D hydrogen chain and the 2-D Hubbard model. The results of the calculations are plotted in Figure~\ref{fig:scaling}. The figure shows the impressive reduction in the CPU time even for a chain of H$_{20}$ which only contains 20 electrons. Further, this speed up increases substantially with the size of the system because of the $N^4$ versus the $N^{1.5}$ scaling of the standard algorithm and the new algorithm respectively. The $N^{1.5}$ scaling of the Hydrogen chain is surprising given the fact that the lowest scaling operation in the Table~\ref{tab:cpu} is $O(N^2)$. The reason for the lower scaling is as follows, the largest CPU time is consumed in calculating the local energy. In calculating the local energy one has to loop over determinants $\M$ in Equation~\ref{eloc:hci}, and we expect $N^2$ scaling because of the double excitation from a doubly occupied orbital $i$ to an empty orbital $a$ with the Hamiltonian transition matrix element equal to $\langle a a| i i\rangle$. Because of the long range nature of the Coulomb interaction this integral decays slowly and we expect it to be important for even very large systems. However, this excitation often does not apply because most determinants $\N$ and $\M$ contain several singly occupied orbitals; the double occupation is suppressed due to the locality of the orbitals. For the hydrogen chain in the minimal basis with a bond length of 2.2 $a_0$, the integrals for the type $\langle ab|ij\rangle$, where $a$,$b$ and $i$,$j$ are spatially close together are almost always very small.

The almost linear scaling of the Hubbard model is due to the fact that only a linear number of local single excitations are present. The scaling of the local energy calculations is thus almost exactly linear, but with the relatively small quadratic scaling expense of updating the determinant, the overall scaling gradually increases with the size of the system and asymptotically will approach quadratic scaling.

\subsection{Performance of the optimizer}\label{sec:optimizer}
\begin{figure}
\begin{center}
\includegraphics[width=0.48\textwidth]{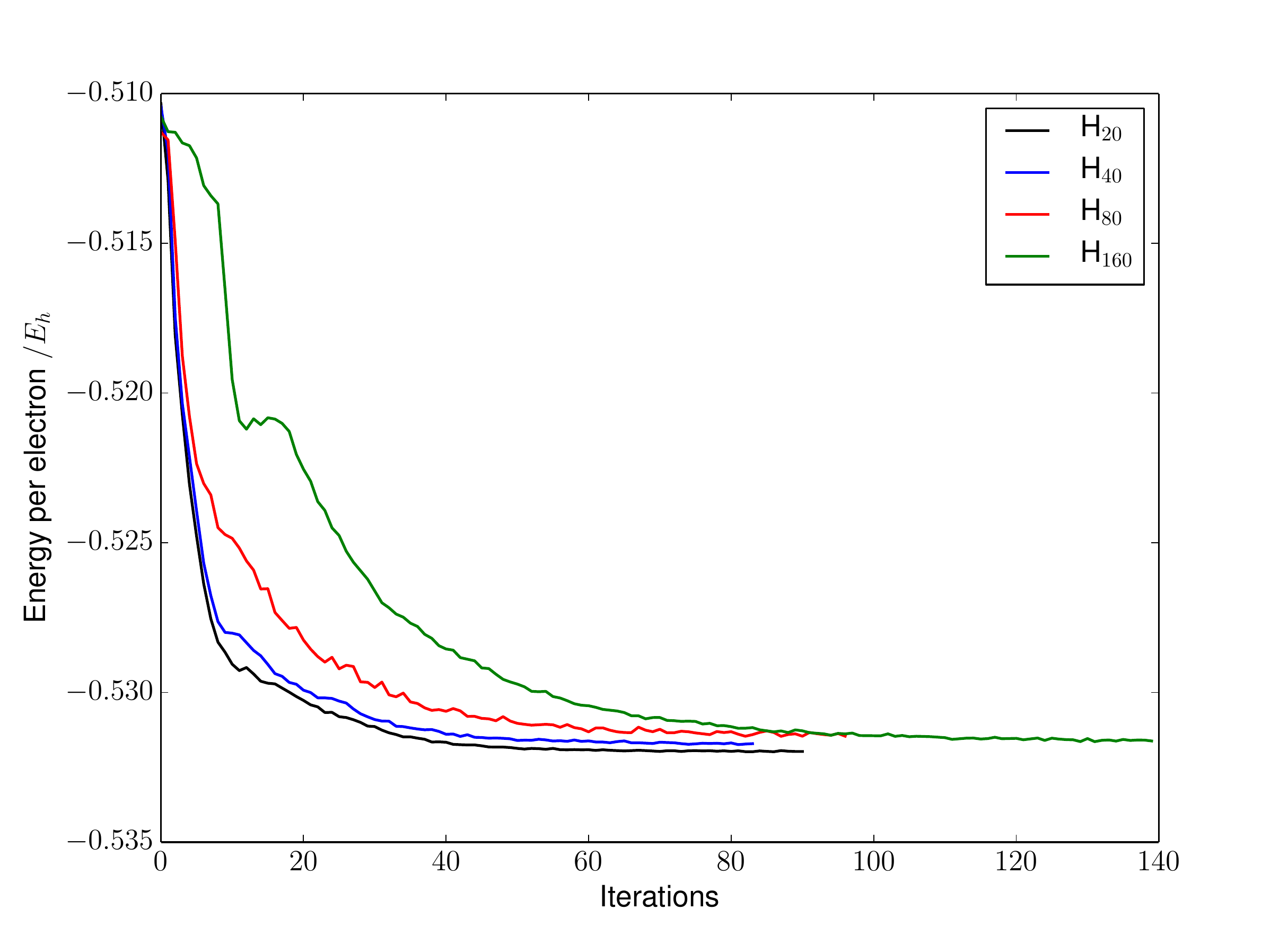}
\end{center}
\caption{The graph shows the energy per electron calculated using the CPS-Slater wavefunction with a 5-site correlator and a unrestricted Hartree Fock determinant as a function of the AMSGrad iterations. Note that in all these calculations the H-H bond length was 2.2 $a_0$, an $\epsilon=10^{-4}$ was used and a system size independent $n_s=240,000$ stochastic samples per optimization step were used.}
\label{fig:solver}
\end{figure}

Figure~\ref{fig:solver} shows the energy per electron as the AMSGrad optimizer is used to minimize the energy of the wavefunction consisting of a product of 5-site CPS and a UHF determinant for H$_{20}$, H$_{40}$ , H$_{80}$ and H$_{160}$ molecules. The number of stochastic samples $n_s$ per optimization step was equal to 240,000 and was the same for all hydrogen chain lengths. It is interesting to note that the rate of convergence of the different hydrogen chains is almost independent of the size of the system. The slightly slower convergence in case of H$_{160}$ and H$_{80}$ was due to the fact that in the initial few iterations we had to use a smaller step size of $\alpha = 0.001$. After performing 5 iterations the step size was increased to $\alpha=0.01$ and was subsequently kept constant. This was necessary to build up reasonable estimates of $\mathbf{n}^{(i)}$ and $\mathbf{m}^{(i)}$, the exponentially decaying moving average of the first and second moment of the gradient (see Equations~\ref{eq:m1} and Equation~\ref{eq:m2}) without which the energy tends to fluctuate wildly for these larger systems. Finally, Table~\ref{tab:hchain} shows that the same energy/electron down to three decimal places is obtained by carrying out the optimization using virtually the same setting in all four cases. This demonstrates that the AMSGrad optimizer is fairly tolerant to the absolute magnitude of the noise and is likely to deliver the same relative accuracy as long as the relative noise is kept constant with the size of the system. %This observation is unlikely to hold for the SR and LM optimizers because as was already mentioned, these optimizers introduce a bias in the final calculated wavefunction that is a non-linear function of the stochastic noise and so we expect it to increase with the system size.

Perhaps the most impressive aspect of the calculation is that the CPU cost of performing the optimization for the H$_{160}$ molecule was merely 25 CPU hours, which when split between 2 nodes containing 24 processors each amounted to a wall time of just over half hour.

\begin{table}
\caption{The table shows the energy per electron for the hydrogen chain of various lengths using the CPS-Slater wavefunction with a 5-site correlator and a unrestricted Hartree Fock determinant. The bond lengths were equal to 2.2 $a_0$ in all cases and the DMRG energy is essentially exact in all these calculations. The stochastic error bars on the VMC energies are less than 0.1 E$_h$, however we expect the error due to incomplete optimization to be on the order of 1 mE$_h$.}\label{tab:hchain}
\begin{tabular}{lcccc}
\hline
\hline
Molecule&~~~~&VMC&~~~&DMRG\\
\hline
H$_{20}$&& -0.531&& -0.532\\
H$_{40}$&& -0.531&& -0.532\\
H$_{80}$&& -0.531&& -0.532\\
H$_{160}$&& -0.531&& -0.532\\
\hline
\end{tabular}
\end{table}

\subsection{2-D Hubbard Model Benchmark}
\begin{figure}
\begin{center}
\includegraphics[width=0.2\textwidth]{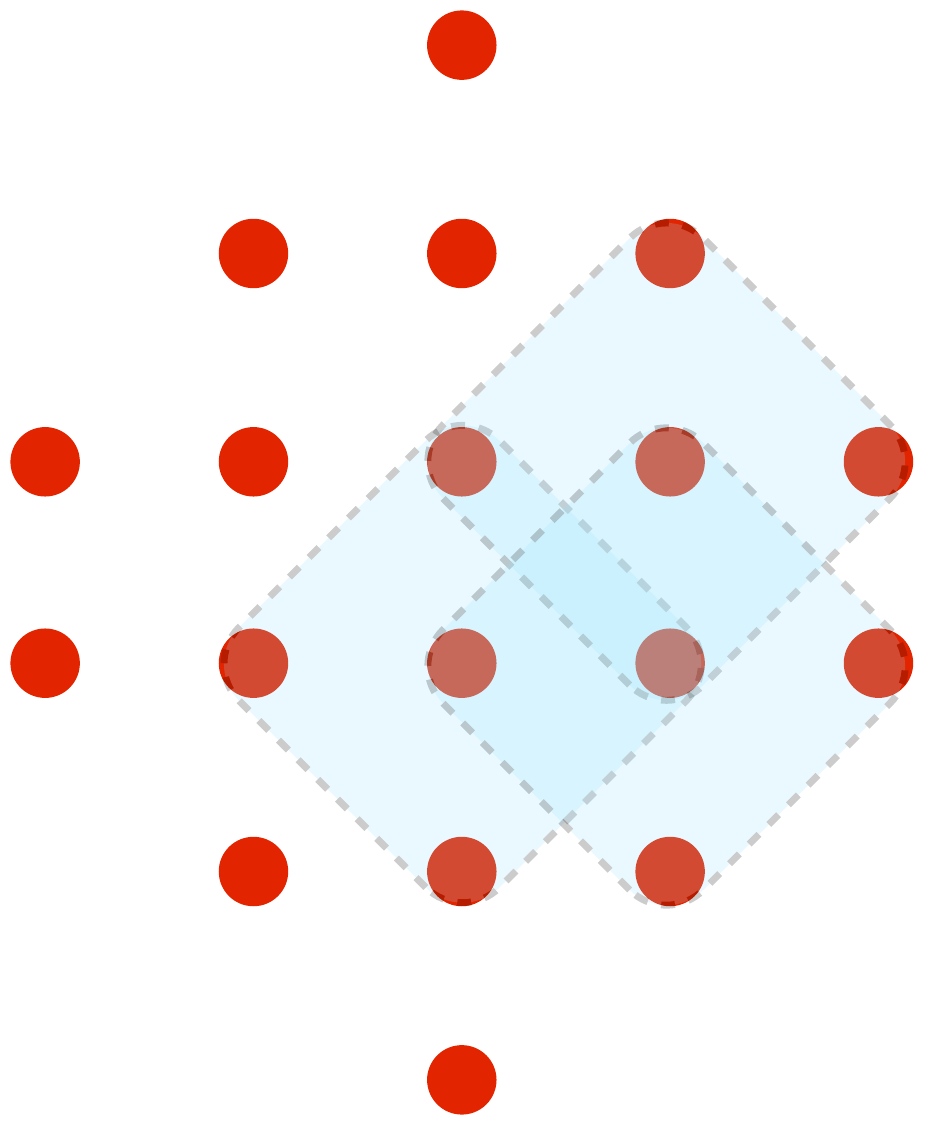}
\end{center}
\caption{The figure shows the 18-site Hubbard model with three overlapping 5-site correlators. The wavefunction consists of such overlapping 5-site correlators, one centered on each site.}
\label{fig:lattice}
\end{figure}

Here we present benchmark results on the 2-D Hubbard model  with periodic boundary conditions that is tilted by 45$^o$ as shown in Figure~\ref{fig:lattice}. The tilted lattice is used because one can obtain a restricted Hartree Fock  solution, which simplifies the subsequent optimization of the CPS-UHF wavefunction. All results are calculated using 5-site overlapping correlators, that are centered on each site of the lattice as shown in Figure~\ref{fig:lattice}. The largest lattice considered here consisted of 162 orbitals and contained $\sim 2\times 10^5$ variational parameters. All optimizations took less than 250 iterations with each iteration using between 7$\times 10^6$-8$\times 10^6$ Monte Carlo samples, embarrassingly parallelized over 100 cpu cores. The overall cost of the largest calculation was about 700 CPU hours which equaled 7 hours of wallclock time. This is quite satisfactory for a quantum Monte Carlo calculation of this size. It is worth pointing out that the converged results that we have obtained for the 98 site Hubbard model with $U/t=8$ is 0.004 /$t$ lower than the one obtained using the RMSprop algorithm\cite{PhysRevLett.118.176403}. This highlights the importance of using aggressive settings in the adapted SGD methods, but more importantly points to the fact that it is often difficult to decide when the optimization has converged. This is because one often observes a relatively long tail at the end of optimization where the energy very gradually decreases with the number of iterations. We have noticed such an effect in our calculations as well and thus it is difficult to estimate the accuracy of the calculations presented here. This provides one motivation for using the second order methods such as LM, which can be used to greatly speed up the convergence when the wavefunction parameters are already close to their optimal values. 
\begin{table}
\caption{The table shows the energy/electron calculated for 2-D Hubbard model. The results at thermodynamic limit (TL) are exact up to the quoted decimal places because these were obtained using the auxilliary-field quantum Monte Carlo method that has no sign problem at half filling\cite{LeBlanc2015,Qin2016}. The stochastic error bars on the VMC energies are less than 0.1 E$_h$, however we expect the error due to incomplete optimization to be on the order of 1 mE$_h$}
\begin{tabular}{lccc}
\hline
\hline
$N$-Sites & ~~$U/t = 2$~~ & ~~$U/t = 4$ ~~& ~~$U/t = 8$~~\\
\hline
18 & -1.319 & -0.947 & -0.524\\
50 & -1.218 & -0.865 & -0.511\\
98 & -1.191 & -0.855 & -0.510\\
162 & -1.180 & -0.853 & -0.510\\
\hline
TL(exact) & -1.176 & -0.860 & -0.525\\ 
\hline
\end{tabular}
\end{table}

\section{Conclusions}
In this work we have introduced several innovations including the efficient screening of integrals, use of continuous time Monte Carlo to sample determinants and finally the use of AMSGrad which displays robust and efficient convergence of the wavefunction in the model systems studied here.

Out of the three innovations introduced above, the first two can be straightforwardly used for any system, however, more work is needed to study the effectiveness of AMSGrad and other such adaptive SGD methods in optimizing the wavefunction of more realistic systems of interest in quantum chemistry. Such work is already underway and we are also exploring a suitable combination of SGD methods in conjunction with direct second order methods; using the former for the bulk of the optimization and switching to the latter at the end of the optimization when the wavefunction is already nearly converged. 

Finally, we are also exploring two different approaches to go beyond the VMC framework and correct the shortcomings of the wavefunction. First, is the use of stochastic perturbation theory\cite{ShaHolUmr-JCTC-17,doi:10.1063/1.5031140,sharma18} that was recently generalized to correct any variational wavefunction. Second, is the use of fixed node orbital-space projector Monte Carlo\cite{PhysRevLett.72.2442,Haaf1995} that is guaranteed to provide variational energies which are at least as good as the VMC energies.

\section{Acknowledgements}
We would like to thank Cyrus Umrigar, Eric Neuscamman and George Booth for several helpful discussions. The funding for this project was provided by the national science foundation through the grant CHE-1800584.
%\bibliographystyle{biochem}
%\bibliography{refs}
\providecommand{\latin}[1]{#1}
\providecommand*\mcitethebibliography{\thebibliography}
\csname @ifundefined\endcsname{endmcitethebibliography}
  {\let\endmcitethebibliography\endthebibliography}{}

\end{document}